\documentclass[journal=cmatex,manuscript=article]{achemso}

\usepackage{amsmath,amssymb}
\usepackage{gensymb}
\usepackage{textcomp}
\usepackage{graphicx}% Include figure files
\DeclareGraphicsExtensions{.eps,.bmp}
\usepackage{chemformula} % Formula subscripts using \ch{}
\usepackage[T1]{fontenc} % Use modern font encodings
\usepackage{bm}% bold math
\usepackage{hyperref}% add hypertext capabilities

\author{William R. Meier}
\email{javamocham@gmail.com}
\affiliation{Materials Science \& Engineering Department, University of Tennessee Knoxville, Knoxville, Tennessee 37996, USA}%

\author{Richa Pokharel Madhogaria}
\affiliation{Materials Science \& Engineering Department, University of Tennessee Knoxville, Knoxville, Tennessee 37996, USA}%

\author{Shirin Mozaffari}
\affiliation{Materials Science \& Engineering Department, University of Tennessee Knoxville, Knoxville, Tennessee 37996, USA}%

\author{Madalynn Marshall}
\affiliation{Neutron Scattering Division, Oak Ridge National Laboratory, Oak Ridge, Tennessee 37831, USA}%

\author{David E. Graf}
\affiliation{National High Magnetic Field Laboratory, Tallahassee, FL, 32310, USA}%

\author{Michael A. McGuire}
\affiliation{Material Science \& Technology Division, Oak Ridge National Laboratory, Oak Ridge, Tennessee 37831, USA}%

\author{Hasitha W. Suriya Arachchige}
\affiliation{Department of Physics \& Astronomy, University of Tennessee Knoxville, Knoxville, Tennessee 37996, USA}%

\author{Caleb L. Allen}
\affiliation{Department of Physics \& Astronomy, University of Tennessee Knoxville, Knoxville, Tennessee 37996, USA}%

\author{Jeremy Driver}
\affiliation{Department of Physics \& Astronomy, University of Tennessee Knoxville, Knoxville, Tennessee 37996, USA}%

\author{Huibo Cao}
\affiliation{Neutron Scattering Division, Oak Ridge National Laboratory, Oak Ridge, Tennessee 37831, USA}%

\author{David Mandrus}
\email{dmandrus@utk.edu}
\affiliation{Department of Physics \& Astronomy, University of Tennessee Knoxville, Knoxville, Tennessee 37996, USA}%
\affiliation{Materials Science \& Engineering Department, University of Tennessee Knoxville, Knoxville, Tennessee 37996, USA}%
\affiliation{Materials Science \& Technology Division, Oak Ridge National Laboratory, Oak Ridge, Tennessee 37831, USA}%

\date{\today}% It is always \today, today,
\title{Tiny Sc allows the chains to rattle: Impact of Lu and Y doping on the charge density wave in ScV$_6$Sn$_6$}% Force line breaks with \\

\abbreviations{CDW}

\begin{document}

%%%%%%%%%%%%%%%%%%%%%%%%%%%%%%%%%%%%%%%%%%%%%%%%%%%%%%%%%%%%%%%%%%%%%
%% The "tocentry" environment can be used to create an entry for the
%% graphical table of contents. It is given here as some journals
%% require that it is printed as part of the abstract page. It will
%% be automatically moved as appropriate.
%%%%%%%%%%%%%%%%%%%%%%%%%%%%%%%%%%%%%%%%%%%%%%%%%%%%%%%%%%%%%%%%%%%%%
\begin{tocentry}

\includegraphics[width=2.986in]{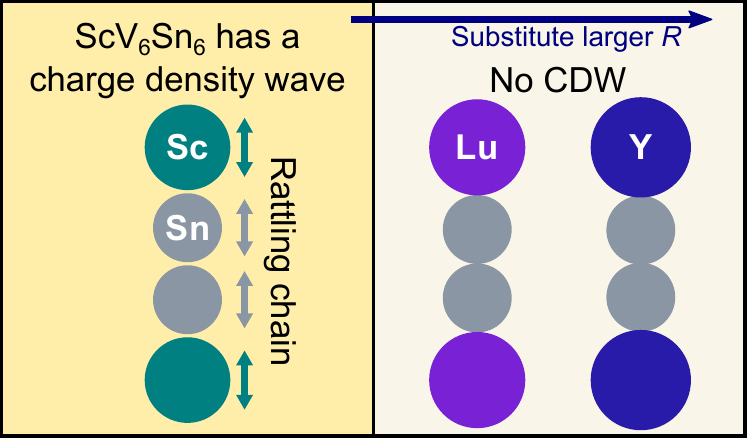}

\end{tocentry}
%% if an abstract is not used by the target journal.
%%%%%%%%%%%%%%%%%%%%%%%%%%%%%%%%%%%%%%%%%%%%%%%%%%%%%%%%%%%%%%%%%%%%%
\begin{abstract}

The kagome metals display an intriguing variety of electronic and magnetic phases arising from the connectivity of atoms on a kagome lattice. 
A growing number of these materials with vanadium kagome nets host charge density waves (CDWs) at low temperatures including ScV$_6$Sn$_6$, CsV$_3$Sb$_5$, and V$_3$Sb$_2$. 
Curiously, only the Sc version of the $R$V$_6$Sn$_6$ HfFe$_6$Ge$_6$-type materials hosts a CDW ($R = $Gd--Lu, Y, Sc). 
In this study we investigate the role of rare earth size in CDW formation in the $R$V$_6$Sn$_6$ compounds. 
Magnetization measurements on our single crystals of (Sc,Lu)V$_6$Sn$_6$ and (Sc,Y)V$_6$Sn$_6$ establish that the CDW is suppressed by substitution of Sc by larger Lu or Y.
Single crystal x-ray diffraction reveals that compressible Sn-Sn bonds accommodate the larger rare earth atoms within loosely packed $R$-Sn-Sn chains without significantly expanding the lattice.
We propose that Sc provides the extra room in these chains crucial to CDW formation in ScV$_6$Sn$_6$.
Our rattling chain model explains why both physical pressure and substitution by larger rare earths hinder CDW formation despite opposite impacts on lattice size. 
We emphasize the cooperative effect of pressure and rare earth size by demonstrating that pressure further suppresses the CDW in a Lu-doped ScV$_6$Sn$_6$ crystal.
Our model not only addresses why a CDW only forms in the $R$V$_6$Sn$_6$ materials with tiny Sc, it also advances to our understanding of why unusual CDWs form in the kagome metals.

\end{abstract}

%This manuscript has been authored by UT-Battelle, LLC under Contract No. DE-AC05-00OR22725 with the U.S. Department of Energy. The United States Government retains and the publisher, by accepting the article for publication, acknowledges that the United States Government retains a non-exclusive, paid-up, irrevocable, world-wide license to publish or reproduce the published form of this manuscript, or allow others to do so, for United States Government purposes. The Department of Energy will provide public access to these results of federally sponsored research in accordance with the DOE Public Access Plan (http://energy.gov/downloads/doe-public-access-plan).
%%%%%%%%%%%%%%%%%%%%%%%%%%%%%%%%%%%%%%%%%%%%%%%%%%%%%%%%%%%%%%%%%%%%%
%% Start the main part of the manuscript here.
%%%%%%%%%%%%%%%%%%%%%%%%%%%%%%%%%%%%%%%%%%%%%%%%%%%%%%%%%%%%%%%%%%%%%
\section{Introduction}
\label{sec:Intro}

A charge density wave (CDW) is an ordered phase of a metallic crystal that appears on cooling. 
It is characterized by both localization of some conduction electrons and an associated atomic displacement which reduce the translational symmetry of the lattice \cite{Zhu2017_MisconceptionsChargeDensityWaves,Gruner1994_DensityWavesInSolids}. 
%Although originally associated with 1D chains \cite{Gruner1994_DensityWavesInSolids}, CDW materials occur across materials with 1D, 2D, and 3D connectivity.[cite many]
CDWs show excellent tune-ability with physical pressure, chemical composition, and disorder \cite{Morosan2006_Superconductivity-Cu-doped-TiSe2,Joe2014_CDWDomainWallsInSCDome-1T-TiSe2,Boubeche2021_SC+CDW-In-I-doped-CuIr2Te4,Kazama2016_Electric+MagProperties-TransitionIn-Sc3TC4,Sangeetha2012_CDWdopedLu2Ir3Si5,Li2023_PressureSuppressCDW-Sm2Ru3Ge5,Veiga2020_QuantumFluctuationNearQCP-Sr3Ir4Sn13-Ca3Ire4Sn13,Goh2015__QuantumCriticalPoint_Ca3Rh4Sn13-Sr3Rh4Sn13,Klintberg2012_QuantumPhaseTransititionSr3Ir4Sn13}. 
In fact, tuning the CDW transition to zero temperature is a good approach to discover new superconductors \cite{Sipos2008_MottToSC-1T-TaS2,Briggs1980_CDW+SC+FermiSurface-NbSe3-Pressure,Zocco2015_CDW+SC-GeTe3-TbTe3-DyTe3,Zeng2021_TiSubstitutionEffectOnSC+CDW-CuIr2Te4,Aulestia2021_PressureEnhancedSC-La2O2Bi3AgS6-withCDW,Eckberg2019_PressureStabilizesSC-Ba1-xSrxNi2As2,Lee2019_SuperconductingCDW-CoDoped-BaNi2As2,Kudo2012_EnhanceSC-CDW-PDoped-BaNi2As2}.

%The 3-4-13 Remeika family [https://doi.org/10.1016/bs.hpcre.2018.10.001] contains a few examples of 3D CDW materials which interact with superconductivity.[cite eg http://dx.doi.org/10.1103/PhysRevB.93.245119 http://dx.doi.org/10.1103/PhysRevLett.109.237008 https://doi.org/10.1016/j.physb.2006.06.037 https://doi.org/10.1088/1742-6596/592/1/012046 https://doi.org/10.1109/TMAG.2013.2255589 https://doi.org/10.1016/j.jallcom.2015.06.198 http://dx.doi.org/10.1103/PhysRevB.88.115113 https://doi.org/10.1103/PhysRevB.88.155122 ] In particular, the evolution of $T_{\mathrm{CDW}}$ and $T_{\mathrm{c}}$ (Sr,Ca)$_{3}$Ir$_{4}$Sn$_{13}$ and (Sr,Ca)$_{3}$Rh$_{4}$Sn$_{13}$ reveal a common phase diagram for physical pressure and across the Sr-Ca isovalent doping series (chemical doping).[http://dx.doi.org/10.1103/PhysRevLett.109.237008, https://doi.org/10.1103/PhysRevB.101.104511] 

\begin{figure}
\includegraphics[width=3.33in]{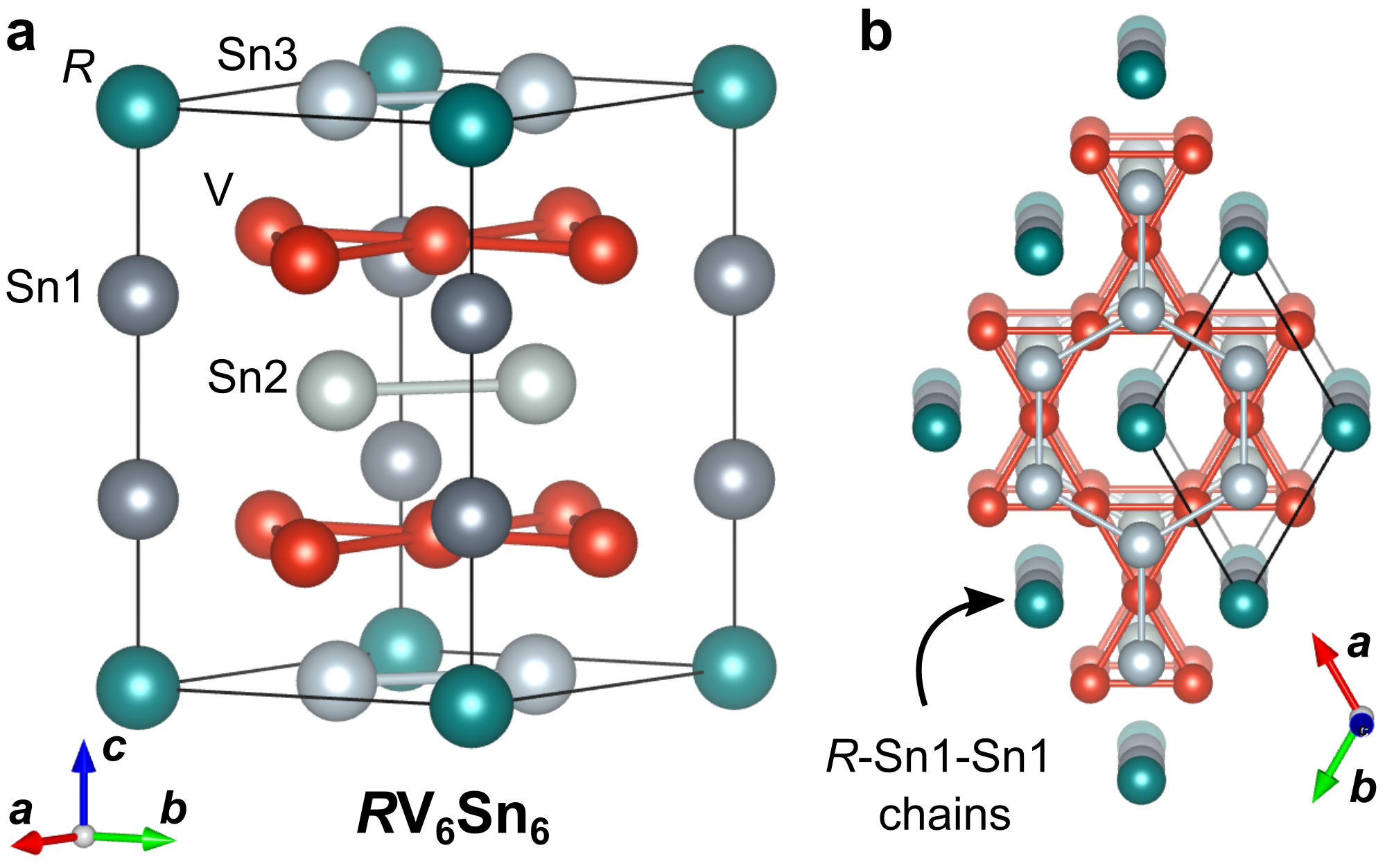}
    \caption{\label{fig:RV6Sn6_Structure} Crystal structure of $R$V$_{6}$Sn$_{6}$. \textbf{a} Stacked vanadium-kagome and Sn2/Sn3-honeycomb sheets along $c$-axis. \textbf{b} $R$-Sn1-Sn1 chains along $\bm{c}$ fill channels in honeycomb and kagome nets. Generated with VESTA\cite{Momma2011_Vesta3}.}    
\end{figure}   

%Colors for R atoms
%Sc 0 128 128 = #008080
%Lu 128 0 255 = #8000FF
%Y 0 0 255 = #0000FF

Kagome metals have recently attracted significant interest for their unusual electronic and magnetic properties arising from kagome sheets of transition metals \cite{Teng2023_CDWinHexFeGe,Meier2020_FlatBandsCoSn-type,Liang2021_3D-CDW+SurfaceDependentVortexCoreStatesKagomeCsV3Sb5,Kang2022_vanHoveSingularities+OriginChargeOrderInCsV3Sb5,Ortiz2021_FermiSurfMapping+NatureOfCDWInCsV3Sb5,Hu2022_NatureOfVanHoveSingularitiesCsV3Sb5,Mozaffari2023_SublinearResistivityVanadiumKagomeMetals-ScV6Sn6-LuV6Sn6,Colman2011_MagAndCrystalStudiesOfGamma-Cu3MgO6H6Cl2,Wang2022_ElectronicPropsAndPhaseTransition-Yb0.5Co3Ge3,Miiller2011_HighlyFrustrateGroundState-YCa3V3O3B4O12,Gui2022_LaIr3Ga2-Superconductor,Allison2022_FeMn3Ge2Sn7O16-KagomeBreaksMagSymmWithPartialSpinOrder,Sales2022_MagOrder-InDoped-CoSn,Thakur2020_AHE-Weyl-Co3Sn2S2,McGuire2021_AFM+LinearMagresist-FeDopedCo3In2S2,Meschke2021_Search+StructFeaturizationMagFrustratedKagome,Jovanovic2022_SimpleChemicalRulesForBandStructuresKagomeMaterials}. 
A growing number of these materials display a CDW at low temperature, especially when vanadium forms the kagome net. 
The $A$V$_3$Sb$_5$ materials ($A$ = K, Rb, Cs) host both superconductivity and curious CDW phases \cite{Ortiz2019_NewKagomePrototypeMaterials-DiscoveryAV3Sb5,Ortiz2021_SuperconductivityInZ2KagomeMetalKV3Sb5,Ortiz2020_CsV3Sb5-Z2TopoKagomeMetalWithSuperconductingGrdState,Liang2021_3D-CDW+SurfaceDependentVortexCoreStatesKagomeCsV3Sb5,Kang2022_vanHoveSingularities+OriginChargeOrderInCsV3Sb5,Ortiz2021_FermiSurfMapping+NatureOfCDWInCsV3Sb5,Hu2022_NatureOfVanHoveSingularitiesCsV3Sb5,Nie2022_CDW-DrivenNematicityInCsV3Sb5,Wang2021_ChiralCDW-CsV3Sb5,Li2021_CDW-WithoutAcosticPhononAnomalyRbV3Sb5+CsV3Sb5,Yu2021_AnomalousHall+CDWCsV3Sb5,Jiang2021_ChiralCDW-KV3Sb5,Yang2020_AnomalousHallInKV3Sb5,Yu2021_CompetitionOfSuperconductivityAndCDWInCsV3Sb5,Du2021_PressureInducedDoubleSuperconductingDomes+ChargeInstabilityKV3Sb5,Oey2022_FermiLevel+DoubleDomeSuperconductivitySnDopedCsV3Sb5,Yin2021_Superconductivity+NormStatePropertiesRbV3Sb5Crystals}. V$_3$Sb$_2$ likely has a CDW transition as well \cite{Wang2022_CDW-like-V3Sb2}.

A CDW has also been identified in the vanadium-kagome metal ScV$_6$Sn$_6$ below 92\,K \cite{Arachchige2022_CDWinScV6Sn6}. 
This is the Sc member of the rare earth $R$V$_6$Sn$_6$ compounds ($R$ = Sc, Y, and Gd--Lu)\cite{Romaka2011_Gd-V-Sn+Er-V-SnSystems+RV6Sn6Phases,Zhang2022_Elec+MagPropsTbV6Sn6-TmV6Sn6,Lee2022_MagPropertiesRV6Sn6}. These adopt the hexagonal HfFe$_6$Ge$_6$ structure type (Fig.~\ref{fig:RV6Sn6_Structure}\textbf{a}) characterized by V-kagome sheets interleaved between Sn2-honeycomb and Sn3-honeycomb sheets. 
Rare earth atoms and Sn1 atoms form a chain that occupy channels in the hexagonal holes in the kagome and honeycomb layers as illustrated in Fig.~\ref{fig:RV6Sn6_Structure}\textbf{b}.
These chains of atoms play a pivotal role in the CDW mode as they displace along the $c$-axis in the modulated structure \cite{Arachchige2022_CDWinScV6Sn6}. 
Several CDW modes appear to compete in ScV$_6$Sn$_6$ \cite{Tan2023_AbundantLatticeInstabilityIn-ScV6Sn6} demonstrated by the strong CDW fluctuations observed above the transition temperature\cite{Cao2023_CompetingCDWInstabilites-ScV6Sn6,Korshunov2023_SofteningOfFlatPhononBand-ScV6Sn6} with impacts on the electrical transport properties \cite{Mozaffari2023_SublinearResistivityVanadiumKagomeMetals-ScV6Sn6-LuV6Sn6}.

It is important to note that the CDW in ScV$_6$Sn$_6$ has a  $\frac{1}{3} \frac{1}{3} \frac{1}{3}$ wave vector \cite{Arachchige2022_CDWinScV6Sn6,Cao2023_CompetingCDWInstabilites-ScV6Sn6,Korshunov2023_SofteningOfFlatPhononBand-ScV6Sn6} in contrast to that observed in the $A$V$_{3}$Sb$_{5}$ materials ($\frac{1}{2} \frac{1}{2} \frac{1}{2}$ or $\frac{1}{2} \frac{1}{2} \frac{1}{4}$)\cite{Ortiz2020_CsV3Sb5-Z2TopoKagomeMetalWithSuperconductingGrdState,Liang2021_3D-CDW+SurfaceDependentVortexCoreStatesKagomeCsV3Sb5,Ortiz2021_FermiSurfMapping+NatureOfCDWInCsV3Sb5,Nie2022_CDW-DrivenNematicityInCsV3Sb5,Wang2021_ChiralCDW-CsV3Sb5,Li2021_CDW-WithoutAcosticPhononAnomalyRbV3Sb5+CsV3Sb5,Neupert2021_ChargeOrder+SuperconductivityAV3Sb5,Mu2021_Superconductivity+CDWInCsV3Sb5FromSb-NQR+V-NMR}. 
Another key difference is that the CDW in $A$V$_{3}$Sb$_{5}$ compounds is dominated by vanadium displacements perpendicular to the hexagonal $c$-axis \cite{Ortiz2021_FermiSurfMapping+NatureOfCDWInCsV3Sb5}.
Numerous investigations of the electronic structure of ScV$_6$Sn$_6$ by photoemission measurements tie the CDW order to the Sn bands instead of the prominent vanadium bands \cite{Cheng2023_ScV6Sn6-STM+ARPES,Kang2023_ARPES+DFT,Hu2023_PhononPromotedCDW-ScV6Sn6_ARPES+Raman+DFT+STM,Lee2023_NatureOfCDW-ScV6Sn6_ARPES+DFT+Phonons+Cr-Dope,Korshunov2023_SofteningOfFlatPhononBand-ScV6Sn6,Hu2023_FlatPhononSoftModes-ScV6Sn6-PhononCalcs}.
There is evidence that the CDW phase has unusual characteristics including anomalous Hall-like responses \cite{Mozaffari2023_SublinearResistivityVanadiumKagomeMetals-ScV6Sn6-LuV6Sn6,Yi2023_AHE-ScV6Sn6} and claims of time reversal symmetry breaking \cite{Guguchia2023_HiddenMagnetismMuSR-ScV6Sn6}. 

So far, none of the other $R$V$_6$Sn$_6$ appear to host CDW order \cite{Lee2022_MagPropertiesRV6Sn6,Mozaffari2023_SublinearResistivityVanadiumKagomeMetals-ScV6Sn6-LuV6Sn6,Pokharel2022_AnisoMag-TbV6Sn6,Pokharel2021_ElectronicProperties-YV6Sn6+GdV6Sn6,Ishikawa2021_GdV6Sn6Properties,Rosenberg2022_FerromagTbV6Sn6,Zhang2022_Elec+MagPropsTbV6Sn6-TmV6Sn6}.
What makes the scandium version special? 
One clue is that the CDW in ScV$_6$Sn$_6$ is suppressed by pressure and disappears before 2.4\,GPa \cite{Zhang2022_ScV6Sn6-pressure}. 
Maybe lattice volume and rare earth size are important.

In this paper, we investigate the impact of rare earth size on CDW formation in the $R$V$_6$Sn$_6$ materials by isovalent substitution of Lu and Y into ScV$_6$Sn$_6$. 
We synthesize single crystals of (Sc$_{1-x}$Lu$_{x}$)V$_6$Sn$_6$ and (Sc$_{1-y}$Y$_{y}$)V$_6$Sn$_6$ and determine the CDW transition temperature with magnetization measurements. 
We find that the CDW phase is suppressed in both cases without significant impact on the lattice size. 
Detailed crystallography reveals that larger $R$ atoms are accommodated by extra room between the Sn1-Sn1 atoms along the $c$-axis (Fig.~\ref{fig:RV6Sn6_Structure}\textbf{a}).
We propose that this space in the loose $R$-Sn1-Sn1 chains is crucial for CDW formation in ScV$_6$Sn$_6$.
Our rattling chain model explains why the CDW is suppressed by pressure as well as by doping with larger rare earths despite their opposite impacts on lattice volume. 
We test our model by confirming that pressure reduces the CDW transition temperature in a sample where instability is already suppressed by Lu-doping.
This study answers why only the Sc version of the $R$V$_6$Sn$_6$ kagome metals hosts CDW order; the small Sc atoms provide extra room for $R$-Sn1-Sn1 chains to rattle and permitting the CDW displacements. 
In addition this system exhibits an deviation from the usual correspondence between chemical and physical pressure.

\section{Experimental}
\label{sec:Experimental}

\subsection{Crystal growth}
\label{sec:Exp_growth}

Crystals of (Sc,Lu)V$_6$Sn$_6$ and (Sc,Y)V$_6$Sn$_6$ were grown from a Sn-rich melt following Arachchige et al. \cite{Arachchige2022_CDWinScV6Sn6} Distilled scandium pieces (Alfa Aesar 99.9\%), distilled Lutetium pieces (Alfa Aesar 99.9\%), yttrium pieces (alfa aesar 99.9\%),  3 mm pieces of vanadium slugs (Alfa Aesar 99.8\%), and tin shot (Alfa Aesar 99.99+\%) were added to 2 or 5\,mL alumina Canfield crucible sets \cite{Canfield2016_FritDiskCrucibles}. An atomic ratio of Sc\,:\,Lu\,:\,V\,:\,Sn or Sc\,:\,Y\,:\,V\,:\,Sn = $1-r$\,:\,$r$\,:\,6\,:\,60 was used for all growths. The crucibles were sealed in silica ampoules filled with about 0.2\,atm argon. These were heated in a box furnace to 1150\,\textdegree C over 12\,h and held for 15\,h to dissolve as much vanadium as possible. Crystals were grown during a 300\,h slow cool to 780\,\textdegree C. The ampoules were then removed from the furnace, inverted into a centrifuge and spun rapidly to fling the remaining liquid away from the crystals.

These growths yielded hard, light-gray metallic hexagonal crystals on crucible walls and vanadium pieces. Mirror-like basal, prismatic and pyramidal facets are common. The Sc and Lu rich crystals were 0.3--3\,mm in size and tend to be blocky with some Sn inclusions. The most Y-rich crystals formed flatter hexagonal plates. 

\subsection{Characterization}
\label{sec:Exp_characterization}

Powder x-ray diffraction (XRD) was carried out using a Bruker D2 Phaser with Cu K$_\alpha$ source and Ni filter for phase identification and to determine lattice parameters.

Single crystal x-ray diffraction measurements at room temperature were carried out using a Rigaku XtaLAB PRO diffractometer. Data collection and integration were done using the Rigaku Oxford Diffraction CrysAlis Pro software \cite{CrysAlisPRO} and the structural refinement was performed using a SHELXTL package \cite{Sheldrick2015_SHELXT-Refine,Sheldrick2015_SHELXT-SpGrp+StructDetermine}. 
Refined structures and .cif files of LuV$_6$Sn$_6$ and YV$_6$Sn$_6$ can be found in the supplemental materials.

Energy disperse spectroscopy (EDS) was performed to estimate the ratios of the rare earth elements. Crystals were mounted in Crystalbond and polished flat. EDS was carried out in a Zeiss EVO scanning electron microscope at 20\,keV.

Magnetization measurements were carried out using a Quantum Design Magnetic Property Measurement System 3 using the Vibrating Sample Magnetometer (VSM) option. 
Crystals were etched in an aqueous 10\,wt\% HCl solution for 12-36\,h to remove surface Sn then attached to a fused silica paddle with GE varnish. 
All measurements presented were measured with the field perpendicular to the hexagonal $c$ axis. 
The fraction of superconducting Sn in each sample was estimated by a 10\,Oe zero field cooled measurement through the transition. 
Estimated beta-Sn fractions ranged from 0.5--7\,vol\%.

To check for superconductivity, resistance measurements down to 0.12 K were performed using the Adiabatic Demagnetization Refrigerator option in a Quantum Design Physical Property Measurement system using silver paste and platinum wire contacts. These results are presented in the supplemental materials.

Two single crystals of (Sc$_{1-x}$Lu$_{x}$)V$_6$Sn$_6$ and one LuV$_6$Sn$_6$ crystal were measured concurrently in the same piston cylinder pressure cell. 
Contacts were made between platinum wire and the samples using Epotek H20E silver epoxy, which was cured at 135\textdegree C for 30\,minutes. 
Daphne 7575 was used as a pressure medium\cite{Stasko2020_DaphneOil7000series} and the value of the pressure was calibrated using the fluorescence of a small ruby chip located near the crystals\cite{Piermarini1975_RubyFluorescenceUnderPressure}. 
The pressure was recorded at room temperature and again at low temperature, by comparing with the values of fluorescence peaks from a ruby sample at ambient pressure. 
Resistance measurements were made for each sample using a Lakeshore 372 resistance bridge with 3708 preamp/scanner. 
The pressure cell was loaded into a Quantum Design Physical Property Measurement System (PPMS) where the temperature was controlled by cooling and warming at a rate of 0.5\,K\,min$^{-1}$ with about 50\,Torr of helium exchange gas in the sample chamber. 
A calibrated Cernox thermometer was fixed to the outside of the pressure cell next to the copper sample wires to accurately determine the temperature of the measured crystals.

\section{\label{sec:Results}Results}
\subsection{\label{sec:Results_Lattice}Lattice trends}

\begin{figure}
\includegraphics[width=3.33in]{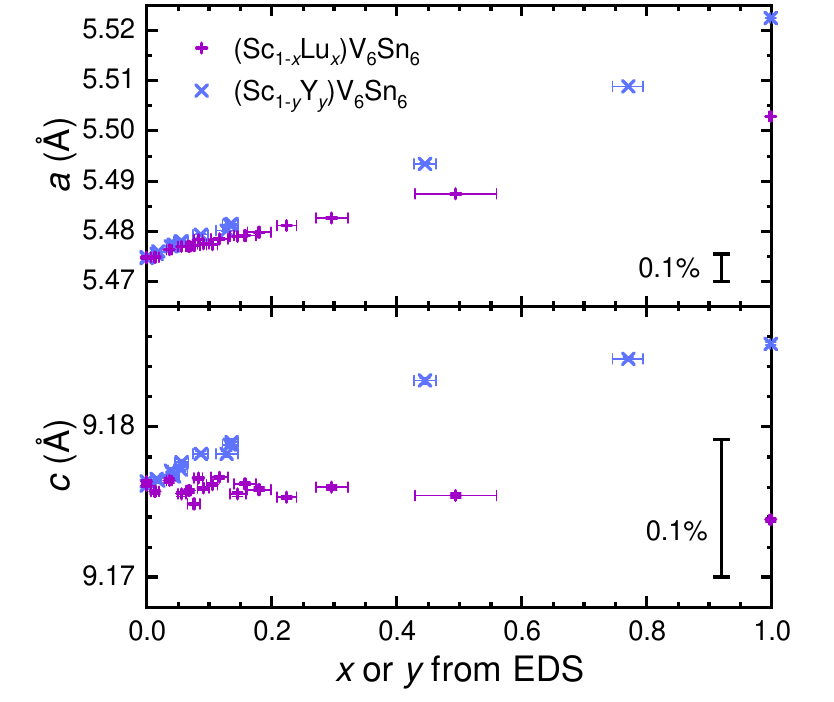}
    \caption{\label{fig:LatticeVsEDS} Variation of lattice parameters for (Sc$_{1-x}$Lu$_{x}$)V$_6$Sn$_6$ and (Sc$_{1-y}$Y$_{y}$)V$_6$Sn$_6$ doping series. Horizontal error bars represent standard deviations of measured $x$ and $y$ by EDS. Vertical error bars are the standard uncertainties from refinements of the lattice parameters. Black vertical bars communicate the size of a 0.1\% change in lattice parameters.}    
\end{figure}   

Figure~\ref{fig:LatticeVsEDS} presents the evolution the lattice parameters as Y and Lu are doped into ScV$_6$Sn$_6$. 
$a$ and $c$ vary continuously across both series indicating that a complete solid solution exists. 
$a$ increases linearly across with a 0.52\% and 0.88\% expansion to LuV$_6$Sn$_6$ and YV$_6$Sn$_6$, respectively. The $c$ parameter increases weakly with Y doping (0.10\%) and subtly decreases for the Lu series (-0.02\%). Ionic radii provide an imperfect proxy of the size $R$ atoms in $R$V$_6$Sn$_6$. The radii of Sc$^{+3}$, Lu$^{+3}$, and Y$^{+3}$ are quoted as 0.87, 0.977, and 1.019 \AA~at 8-coordinated sites in oxide materials \cite{Shannon1976_Radii}. Considering the Lu and Y ions are 12\% and 17\% larger than Sc, it is surprising that the changes in the $c$ lattice parameter are so weak across the substitution series. 

\subsection{\label{sec:Results_Magnetization}Magnetization}

\begin{figure}
\includegraphics[width=3.33in]{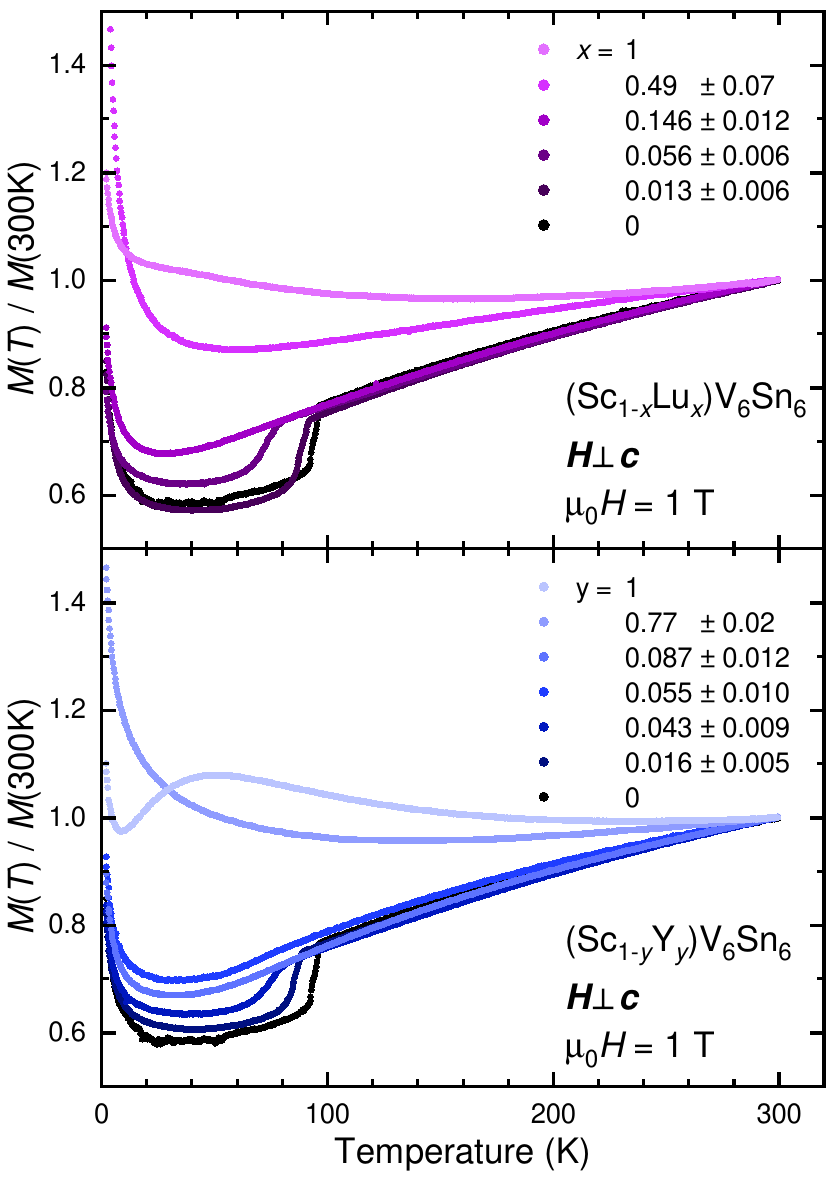}
    \caption{\label{fig:MT} Normalized magnetization vs temperature curves for Lu and Y doped ScV$_6$Sn$_6$. All curves presented were measured on warming after cooling in zero field. } 
\end{figure}   

Next we will explore how Lu and Y substitution impacts the CDW in ScV$_6$Sn$_6$. 
Figure~\ref{fig:MT} presents the normalized magnetization curves for selected Lu and Y doped samples. 
The magnitude of the susceptibility at room temperature is comparable for all samples measured, ranging from 7.5--9.0\,$\times 10^{-7}$ emu\,g$^{-1}$\,Oe$^{-1}$ (6.2--8.2\,$\times 10^{-5}$\,cm$^3$\,(mol atom)$^{-1}$) consistent with Pauli paramagnetism \cite{Kittel2004_SolidStatePhysics}. 
The CDW transition in ScV$_6$Sn$_6$ manifests as a sharp drop of susceptibility on cooling through the first order transition\cite{Arachchige2022_CDWinScV6Sn6}. 
This signature begins at 94\,K (black) and shifts to lower temperatures with increasing Lu or Y content. 
At the same time, the size of drop decreases. 
The broadening of the transition step is likely due to chemical inhomogeneity observed by EDS.

\subsection{\label{sec:Results_PhaseDiagrams}Phase Diagrams}

\begin{figure}    \includegraphics[width=3.33in]{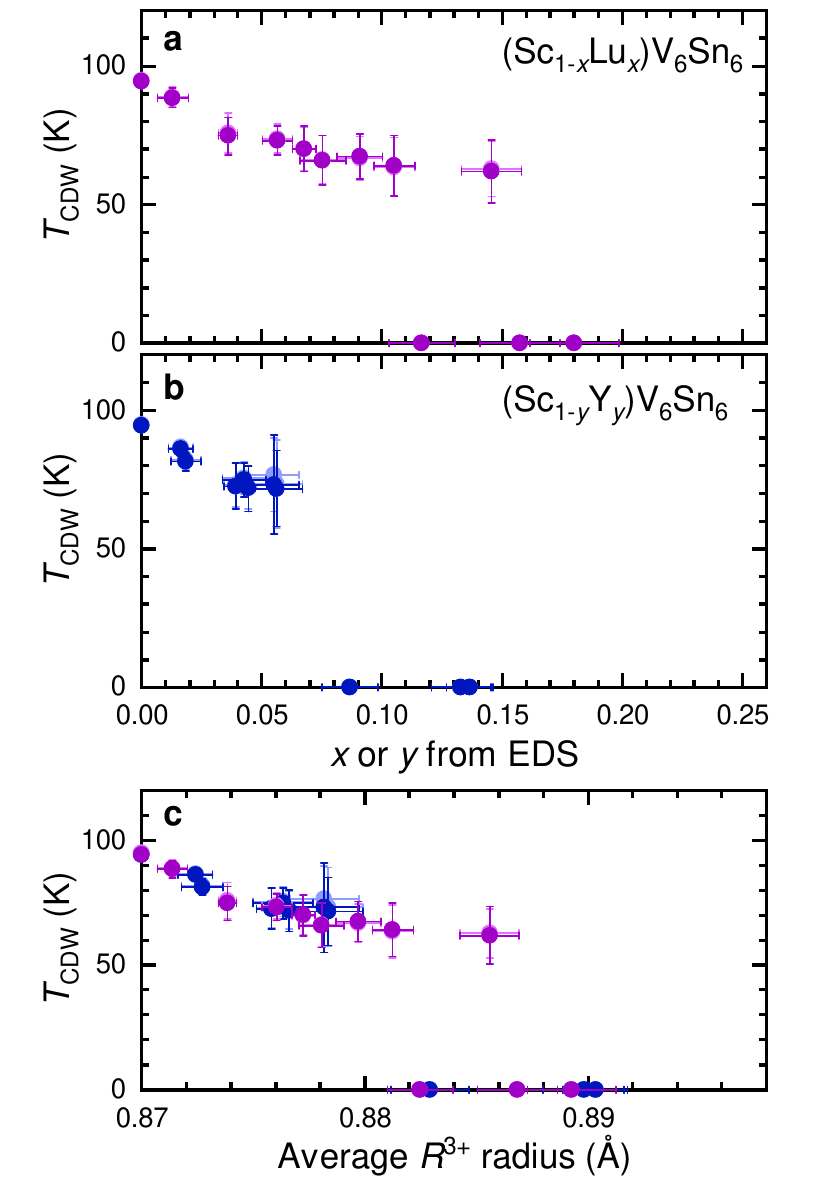}
    \caption{\label{fig:DopingPhaseDiagram} Evolution of $T_\mathrm{CDW}$ with Lu and Y doping of ScV$_6$Sn$_6$.
    \textbf{a} and \textbf{b} Charge density wave temperature ($T_\mathrm{CDW}$) vs composition. Horizontal error bars represent standard deviations of Lu/(Sc+Lu) and Y/(Sc+Y) determined by EDS. 
    Vertical bars represent the estimated transition width of the transition based on the drop in the $M(T)$ measurements.
    Light and dark symbols represent the transition observed on warming and cooling respectively. \textbf{c} $T_\mathrm{CDW}$ vs average ionic radius of $R$ atom\cite{Shannon1976_Radii}.}    
\end{figure}   

Figure~\ref{fig:DopingPhaseDiagram}\textbf{a} and \textbf{b} summarizes suppression of the CDW phase with Lu and Y substitution extracted from magnetization. $T_\mathrm{CDW}$ falls smoothly then discontinuously around a critical doping ($x\approx 0.13$ and $y\approx 0.07$) where we lose the step-like signature. This is not the usual behavior we expect for a quantum critical point were the transition temperature is smoothly reduced to 0\,K.\cite{Coleman2005_QuantumCriticalityReview} Instead, our CDW transition is first order so the phase can disappear more abruptly. No new transitions are observed at higher values of $x$ and $y$.

It is important to note that less Y is required to destroy the CDW than Lu. 
This might arise from yttrium's larger size.  Figure~\ref{fig:DopingPhaseDiagram}\textbf{c} shows that the trend of $T_\mathrm{CDW}$ vs average $R^{3+}$ ionic radius has similar behavior for the Lu an Y series. 
The Lu and Y are also heavier atoms than Sc (174.97, 88.906, and 44.956\,g mol$^{-1}$, respectively). This could be playing a role in suppressing the CDW as proposed by Hu et al. \cite{Hu2023_FlatPhononSoftModes-ScV6Sn6-PhononCalcs}. 
We suggest that this is less important than size because Lu is heavier than Y and we observe that less Y is needed to destabilize the CDW. 

Suppression of a CDW can lead to superconductivity. ADR resistivity measurements revealed no evidence of superconductivity down to at least 150\,mK in Lu doped samples with $x=0.157\pm0.017$ and $0.180\pm0.019$ as well as Y doped samples with $y=0.055\pm0.010$ and $0.087\pm0.012$.

%\subsection{\label{sec:Results_Superconductivity}Superconductivity?}

\section{\label{sec:Discussion}Discussion}

%Message: ScV6Sn6 has a CDW and Lu and Y do not because the small Sc atoms leave extra room in the R-Sn1 displacements of the CDW mode.
		
To simplify our understanding of the evolution of phase transitions we often consider that chemical and mechanical pressure have the same impact on the electronic and lattice subsystems. 
%In this picture, substituting an atom with a smaller dopant is considered as positive physical pressure as the lattice shrinks as it would under load. Conversely, substituting with a larger atom is expected to be like applying negative mechanical pressure. 
If lattice volume has a strong impact on the stability of a phase, then we might expect the transition temperature to rise as we expand the lattice with doping (negative chemical pressure) and fall as we compress the lattice mechanically or through doping.

This correspondence between chemical and physical pressure is epitomized by the CDW and superconducting 3-4-13 Remeika phases. 
Specifically, the (Sr,Ca)$_{3}$Ir$_{4}$Sn$_{13}$ and (Sr,Ca)$_{3}$Rh$_{4}$Sn$_{13}$ substitution series (chemical pressure) exhibit common evolution of the CDW and superconducting critical temperatures with physical pressure\cite{Klintberg2012_QuantumPhaseTransititionSr3Ir4Sn13,Goh2015__QuantumCriticalPoint_Ca3Rh4Sn13-Sr3Rh4Sn13}.

Our investigation of the CDW evolution in isovalent doped ScV$_6$Sn$_6$ reveals behavior that deviates from the universal chemical-physical pressure picture. 
We observe that the $T_\mathrm{CDW}$ decrease with negative chemical pressure as we substitute in the larger Lu and Y atom (Fig.~\ref{fig:DopingPhaseDiagram}). 
Zhang et al. reveal that physical pressure also reduces $T_\mathrm{CDW}$ \cite{Zhang2022_ScV6Sn6-pressure}. The observation that CDW in ScV$_6$Sn$_6$ is suppressed in both cases suggests a new model is needed to explain this behavior.        
  
\begin{figure}
    \includegraphics[width=3.33in]{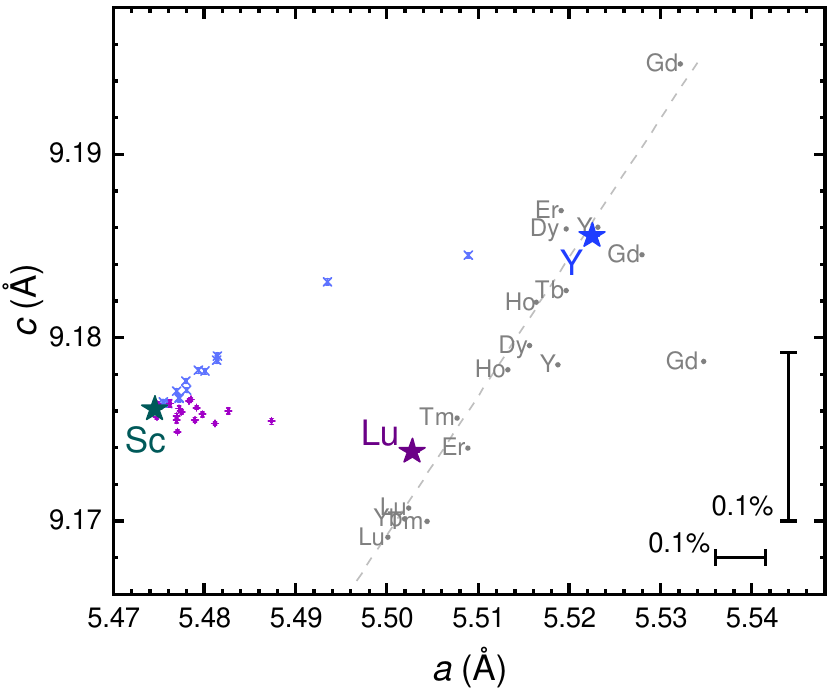}
    \caption{\label{fig:cVsa} Comparison of room temperature $c$ and $a$ lattice parameters for doped ScV$_6$Sn$_6$ and other reported $R$V$_6$Sn$_6$. Gray dots label reported lattice parameters \cite{Lee2022_MagPropertiesRV6Sn6,Pokharel2021_ElectronicProperties-YV6Sn6+GdV6Sn6,Guo2022_TriangularKondo+QCP-YbV6Sn6,Romaka2011_Gd-V-Sn+Er-V-SnSystems+RV6Sn6Phases}. Stars represent the lattice parameters of ScV$_6$Sn$_6$, LuV$_6$Sn$_6$, and YV$_6$Sn$_6$ determined by powder XRD. The small purple and blue dots are the lattice parameters of the (Sc,Lu)V$_6$Sn$_6$ and (Sc,Y)V$_6$Sn$_6$ doping series, respectively.}
\end{figure}  

Let's begin by looking at how the rare earth atom size actually impacts the lattice.
Figure~\ref{fig:cVsa} illustrates how the $c$ and $a$ lattice parameters of $R$V$_6$Sn$_6$ compounds depends on $R$ at room temperature. 
First, observe that most of the the lattice parameters of $R$V$_6$Sn$_6$ compounds (gray points) lie on a linear trend (dashed line) with the smaller, late rare earths with smaller $c$ and $a$. 
Sc is the smallest rare earth atom. If only the size matters, we might expect the lattice parameters of ScV$_6$Sn$_6$ (dark green star) to lie on the lower left of this trend. 
This is not the case.
It has a smaller $a$ but a comparable $c$ to the Lu compound (purple star).
Obviously, the lattice size is not simply controlled by the size of the rare earth atom.
Our doping series emphasize this anomalous lattice evolution in Fig.~\ref{fig:cVsa} (small blue and purple symbols) as the lattice parameters interpolate between the end members. 
%Note that blue (Sc,Y)V$_6$Sn$_6$ series has as weak deviation from Vegard's law [citation]. 

\begin{figure}
    \includegraphics[width=3.33in]{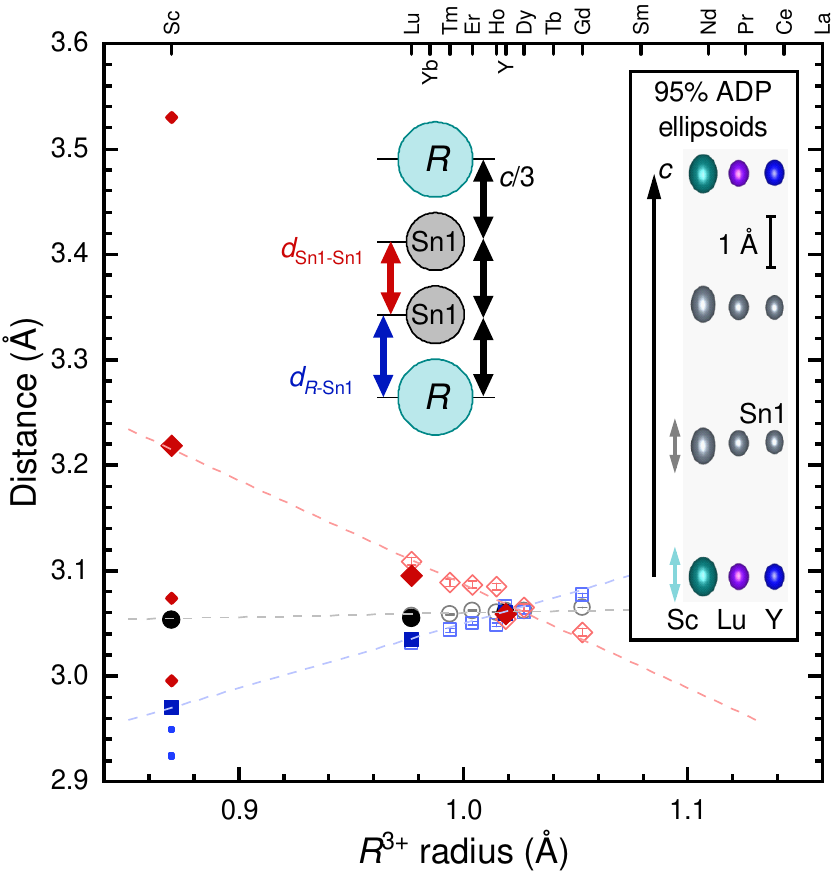}
    \caption{\label{fig:BondVsRadius} Evolution of bond distances with $R^{3+}$ rare earth ionic radii (coordination number = 8 from Shannon et al.\cite{Shannon1976_Radii}). 
    Large, pale colored marks come from literature \cite{Romaka2011_Gd-V-Sn+Er-V-SnSystems+RV6Sn6Phases}.
    %[https://doi.org/10.1016/j.jallcom.2011.06.095 ]. 
    Large solid symbols represent distances determined by single crystal x-ray diffraction refinements in this study and Arachchige et al. \cite{Arachchige2022_CDWinScV6Sn6}. 
    Tiny symbols present the Sn1-Sn1 (red) and $R$-Sn1 (blue) bond lengths in the CDW phase of ScV$_6$Sn$_6$ \cite{Arachchige2022_CDWinScV6Sn6}. 
    The boxed inset depicts the 95\% displacement ellipsoids of the $R$ and Sn1 atoms in the refined structures of the Sc, Lu, and Y compounds. The atomic displacement parameters are plotted in supplemental materials.}
\end{figure}  

We have established that the $c$ lattice parameter has a nontrivial dependence on the rare earth ion. 
How are larger $R$ atoms accommodated in the crystal structure if $c$ is only weakly affected?
Figure~\ref{fig:BondVsRadius} presents the evolution of key bond distances with $R$ ionic radius across the $R$V$_6$Sn$_6$ materials determined by single crystal XRD. 
To begin, note that larger rare earth atoms produce a larger $R$-Sn1 bond distance (blue) but the $c$ lattice parameter (black, plotted as $\frac{c}{3}$) increases far more slowly.
This requires that the Sn1-Sn1 distance must decrease (red). 
In other words, larger rare earths compress the Sn1-Sn1 bond with little impact on the unit cell height. 
Apparently, there is extra room in the $R$-Sn1-Sn1 chain to accommodate larger atoms.

\begin{figure}
    \includegraphics[width=3.33in]{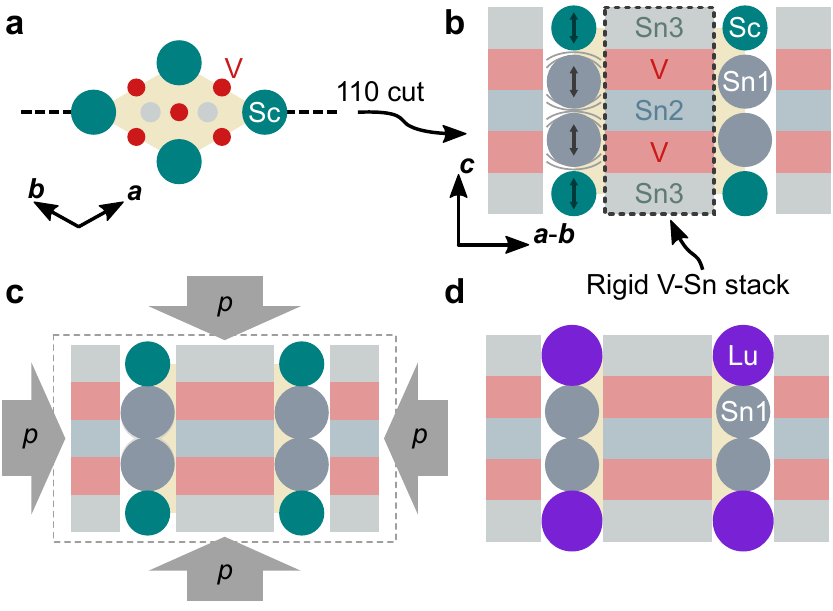}
    \caption{\label{fig:StackCartoons} Cartoon of $R$-Sn1-Sn1 stack. \textbf{a} view along [001] of ScV$_6$Sn$_6$ structure. \textbf{b} (110) cut through the ScV$_6$Sn$_6$ structure depicting rigid stack of V-kagome and Sn-honeycomb sheets (colored rectangles) and loosely packed chain of Sc and Sn1 atoms providing room for them to rattle along $\bm{c}$. \textbf{c} applying pressure compresses the lattice and removes the extra space in the Sc-Sn1-Sn1 chains. \textbf{d} replacing Sc with larger Lu compresses the $R$-Sn1-Sn1 chain and removes the room for the atoms to rattle.}
\end{figure}   

What is setting the size of $c$ if the $R$-Sn1-Sn1 chain does not?
We propose that the stack of V-kagome and Sn2/Sn3-honeycomb sheets do. 
Start by considering the $R$V$_6$Sn$_6$ as two subsystems; $R$-Sn1-Sn1 chains filling tubes in a superstructure of stacked V-kagome and Sn-honeycomb sheets (illustrated in Fig.~\ref{fig:RV6Sn6_Structure}\textbf{b}). 
Figure~\ref{fig:StackCartoons} depicts a cut along the (110) plane presenting the sections of V and Sn sheets as colored rectangles.

ScV$_6$Sn$_6$ and  LuV$_6$Sn$_6$ have nearly identical $c$ values suggesting that the stacked Sn and V sheets are setting $c$ and small Sc atom leave the channels under-filled (Fig.~\ref{fig:StackCartoons}\textbf{b}).  
This facilitates low-energy rattling displacements of the Sn1. 
This rattling is evident in the atomic displacement parameters (ADPs) from our single crystal XRD refinements (see supplemental materials). 
The $c$-axis displacement parameters, $U_{33}$, for the $R$ and Sn1 atoms are twice as large in ScV$_6$Sn$_6$ than in the Lu or Y compounds. \textcolor{red}{}
We illustrate this difference in the inset of Fig.~\ref{fig:BondVsRadius} by depicting the 95\% displacement ellipsoids for the $R$ and Sn1 atoms. 
The Sc and Sn1 atoms in the first column show taller ellipsoids representing larger position deviations in this direction. 
This reflects the strong CDW fluctuation observed in ScV$_6$Sn$_6$ by inelastic and diffuse x-ray scattering \cite{Cao2023_CompetingCDWInstabilites-ScV6Sn6,Korshunov2023_SofteningOfFlatPhononBand-ScV6Sn6}.

These are precisely the displacements that are most prominent in ScV$_6$Sn$_6$'s CDW phase. 
The small symbols in Fig.~\ref{fig:BondVsRadius} are the bond lengths in the refined CDW structure of ScV$_6$Sn$_6$ from Arachchige et al. \cite{Arachchige2022_CDWinScV6Sn6}. 
The Sc-Sn1 bonds shrink a little (small blue dots) but, the Sn1-Sn1 bonds vary more dramatically (small red dots). 
Two Sn1-Sn1 distances shorten by 4.5\% and 6.9\% while the final third of bonds grow by 9.7\%. 
These Sn1-Sn1 bonds are key feature of CDW and modify the Sn1 $p_z$ orbital states \cite{Korshunov2023_SofteningOfFlatPhononBand-ScV6Sn6,Hu2023_FlatPhononSoftModes-ScV6Sn6-PhononCalcs}.

We propose that the CDW instability needs the extra space in $R$-Sn1-Sn1 for the modulations of Sn1-Sn1 distances observed in the CDW. 
If the column is packed too tightly by larger $R$ atoms or physical pressure, the displacements are penalized. 

Next, we will examine how we might expect modifications of ScV$_6$Sn$_6$ to impact loose chains of $R$-Sn1-Sn atoms and, the CDW this facilitates. 
Applying pressure suppresses $T_\mathrm{CDW}$ to 0\,K by 2.4\,GPa \cite{Zhang2022_ScV6Sn6-pressure}. 
Pressure compresses the stack of V and Sn sheets (Fig.~\ref{fig:StackCartoons}\textbf{c}).
This constrains the chain of atoms which shortens the Sn1-Sn1 bonds. 
The CDW is penalized as the extra room in the chain is removed.

Doping ScV$_6$Sn$_6$ with Lu and Y quickly kills the CDW. 
In this case, introducing larger atoms fills up the extra room in the $R$-Sn1-Sn1 chains (Fig.~\ref{fig:StackCartoons}\textbf{d}), compressing the Sn1-Sn1 bonds and suppressing the CDW instability. 
This explains not only why Lu and Y doping suppress the CDW but also why no CDW transition is observed in $R$V$_6$Sn$_6$ with $R$ bigger than Sc. 

\begin{figure}
\includegraphics[width=3.33in]{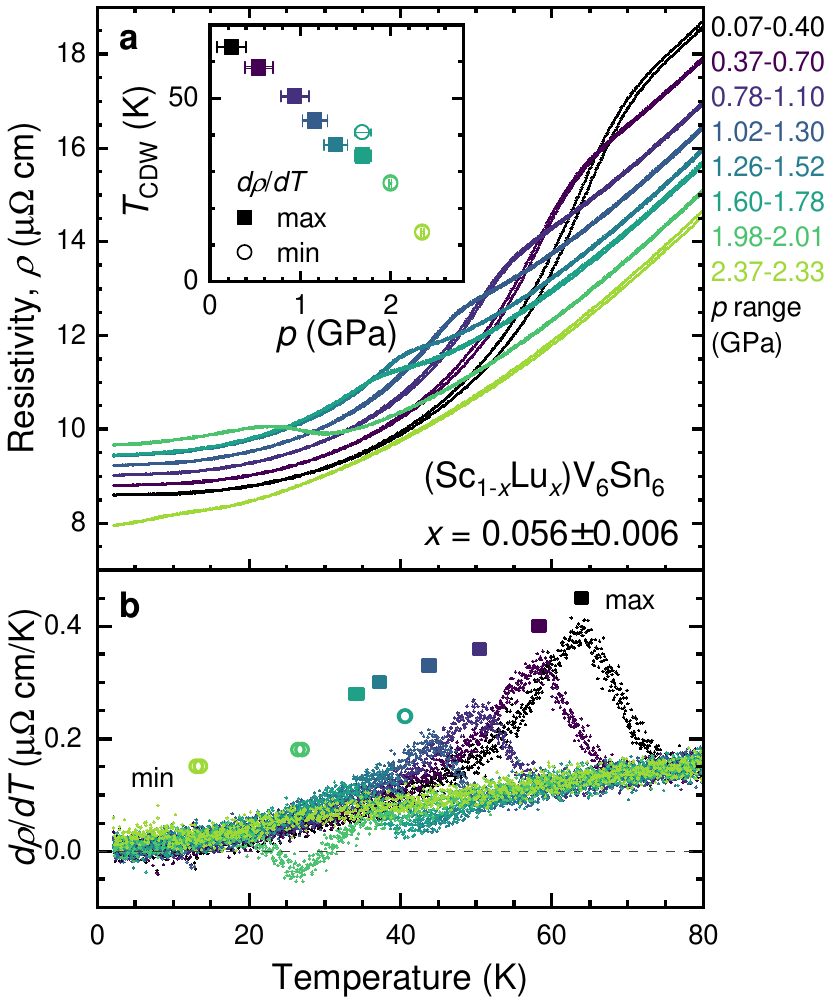}
    \caption{\label{fig:LuDopedPressure} Pressure suppresses the CDW in Lu-doped ScV$_6$Sn$_6$. \textbf{a} Resistivity vs temperate plots for (Sc$_{0.944}$Lu$_{0.056}$)V$_6$Sn$_6$ at increasing pressures, $p$. Resistivity was measured with current perpendicular to the $c$-axis. The pressures at low and high temp are listed on the right. The inset shows pressure dependence of $T_\mathrm{CDW}$. \textbf{b} Derivative of $\rho(T)$ highlighting maxima (squares) and minima (circles) used to define $T_\mathrm{CDW}$.}    
\end{figure}   

This rattling chain mechanism is distinct from the familiar correspondence between chemical and physical pressure.
The CDW in ScV$_6$Sn$_6$ is suppressed by both lattice compression (physical pressure) and expansion by Lu/Y doping. Our model predicts that 
pressure and doping by larger rare earth should cooperate to reduce $T_\mathrm{CDW}$.

We test this prediction by monitoring the evolution of transition temperature with pressure in a crystal where the CDW is already partly suppressed by Lu-doping.
Figure~\ref{fig:LuDopedPressure} presents the resistance vs temperature of (Sc$_{0.944}$Lu$_{0.056}$)V$_6$Sn$_6$ at a series of pressures. 
Near atmospheric pressure (black) the CDW transition is observed as a hysteretic drop in resistance on cooling at roughly 64\,K. 
This drop-like feature decreases in temperature as pressure is increased until it changes character to a jump up on cooling for the highest two pressures (observed in pure ScV$_6$Sn$_6$ under pressure too \cite{Zhang2022_ScV6Sn6-pressure}). 
Applying pressure compresses the V-Sn stack around the already compacted $R$-Sn1-Sn1 chain in the Lu-doped sample further penalizing the CDW displacements.
This observation discredits the possibility that ScV$_6$Sn$_6$ has an optimal lattice volume for CDW formation. In that case applying pressure to the Lu-doped sample should re-stabilize the CDW and increase $T_\mathrm{CDW}$, but this is not what we observe (see supplemental materials).

\section{Conclusions}
\label{sec:Discussion_Conclusions}

In conclusion, we explored how rare earth size impacts CDW formation in the $R$V$_6$Sn$_6$ family. 
Magnetization measurements on single crystals of (Sc,Lu)V$_6$Sn$_6$ and (Sc,Y)V$_6$Sn$_6$ revealed that the CDW transition temperature is reduced with increasing Lu and Y content. 
X-ray diffraction reveals enlightening lattice and bond-length trends. 
These observations and the evolution of the CDW under pressure motivate us to propose a rattling chain model for CDW formation in the $R$V$_6$Sn$_6$ compounds.
We assert that a CDW is only observed in the Sc version of the $R$V$_6$Sn$_6$ compounds because the small Sc atom provide space for Sc and Sn1 atoms to rattle and permit the Sn1-Sn1 bond length modulation observed in the low temperature CDW phase. 
This model explains why both physical pressure and substitution of Sc by larger rare earth atoms suppresses CDW order. This is a curious variation to the usual corresponding effects of chemical and physical pressure on phase stability.
In addition to explaining why ScV$_6$Sn$_6$ is special, our observations and model provide an important leap forward in our understanding of CDWs in the intriguing kagome metals.

\begin{acknowledgement}
\label{sec:Acknowledgment}

We thank Bryan C. Chakoumakos of Oak Ridge National Laboratory for his helpful discussions with the single crystal x-ray refinements.
W.R.M., H.W.S.A., C.L.A., J.D., and D.M. acknowledge support from the Gordon and Betty Moore Foundation’s EPiQS Initiative, Grant GBMF9069. 
S.M. and R.P.M. acknowledge the support from AFOSR MURI (Novel Light-Matter Interactions in Topologically Non-Trivial Weyl Semimetal Structures and Systems), grant\# FA9550-20-1-0322. 
M.M. and H.B.C. were supported by the U.S. Department of Energy (DOE), Office of Science, Office of Basic Energy Sciences, Early Career Research Program Award KC0402020, under Contract DE-AC05-00OR22725. 
M.A.M. acknowledges support for adiabatic demagnetization refrigerator resistance measurements from the US Department of Energy, Office of Science, Basic Energy Sciences, Materials Sciences and Engineering Division. 
A portion of this research used resources at the Spallation Neutron Source, a DOE Office of Science User Facility operated by the Oak Ridge National Laboratory. 
A portion of this work was performed at the National High Magnetic Field Laboratory, which is supported by National Science Foundation Cooperative Agreement No. DMR-2128556 and the State of Florida.

\end{acknowledgement}

%%%%%%%%%%%%%%%%%%%%%%%%%%%%%%%%%%%%%%%%%%%%%%%%%%%%%%%%%%%%%%%%%%%%%
%% The same is true for Supporting Information, which should use the
%% suppinfo environment.
%%%%%%%%%%%%%%%%%%%%%%%%%%%%%%%%%%%%%%%%%%%%%%%%%%%%%%%%%%%%%%%%%%%%%
\begin{suppinfo}
\label{sec:suppinfo}

%A listing of the contents of each file supplied as Supporting Information should be included. For instructions on what should be included in the Supporting Information as well as how to prepare this material for publications, refer to the journal's Instructions for Authors.

The following files are available free of charge.
\begin{itemize}
	\item Supplementary information file including refined structures of LuV$_6$Sn$_6$ and YV$_6$Sn$_6$ as well resistance under pressure measurements for LuV$_6$Sn$_6$ and (Sc$_{1-x}$Lu$_{x}$)V$_6$Sn$_6$ with $x = 0.180\pm0.019$. Adiabatic demagnetization resistance measurements of doped ScV$_6$Sn$_6$ and LuV$_6$Sn$_6$ crystals are included as well.
	\item LuV6Sn6\textunderscore20230509.cif room temperature single crystal refinement of LuV$_6$Sn$_6$
	\item YV6Sn6\textunderscore20230519.cif room temperature single crystal refinement of YV$_6$Sn$_6$
\end{itemize}

\end{suppinfo}

%%%%%%%%%%%%%%%%%%%%%%%%%%%%%%%%%%%%%%%%%%%%%%%%%%%%%%%%%%%%%%%%%%%%%
%% The appropriate \bibliography command should be placed here.
%% Notice that the class file automatically sets \bibliographystyle
%% and also names the section correctly.
%%%%%%%%%%%%%%%%%%%%%%%%%%%%%%%%%%%%%%%%%%%%%%%%%%%%%%%%%%%%%%%%%%%%%

%\bibliography{AllRefrences.bib}% Produces the bibliography via BibTeX.
\providecommand{\latin}[1]{#1}
\makeatletter
\providecommand{\doi}
{\begingroup\let\do\@makeother\dospecials
	\catcode`\{=1 \catcode`\}=2 \doi@aux}
\providecommand{\doi@aux}[1]{\endgroup\texttt{#1}}
\makeatother
\providecommand*\mcitethebibliography{\thebibliography}
\csname @ifundefined\endcsname{endmcitethebibliography}
{\let\endmcitethebibliography\endthebibliography}{}

\end{document}